\documentclass[10pt]{article}
\usepackage{multicol}
\usepackage{graphicx}
\usepackage{amsmath}
\usepackage[a4paper]{geometry}
\usepackage{CJK}

\begin{document}

\begin{center}
{\large\bf Light particle and quark chemical potentials from
negatively to positively charged particle yield ratios corrected
by removing strong and weak decays}

\vskip0.75cm

Hai-Ling Lao$^1$, Ya-Qin Gao$^2$, Fu-Hu
Liu$^{1,}${\footnote{E-mail: fuhuliu@163.com; fuhuliu@sxu.edu.cn}}

\vskip0.2cm

{\small\it $^1$Institute of Theoretical Physics and State Key
Laboratory of Quantum Optics and Quantum Optics Devices,\\ Shanxi
University, Taiyuan, Shanxi 030006, China

$^2$Department of Physics, Taiyuan University of Science and
Technology, Taiyuan, Shanxi 030024, China}

\end{center}

\vskip0.5cm

{\bf Abstract:} The yield ratios of negatively to positively
charged pions ($\pi^-/\pi^+$), negatively to positively charged
kaons ($K^-/K^+$), and anti-protons to protons ($\bar p/p$)
produced in mid-rapidity interval in central gold-gold (Au-Au)
collisions, central lead-lead (Pb-Pb) collisions, and inelastic
(INEL) or non-single-diffractive (NSD) proton-proton ($pp$)
collisions, as well as in forward rapidity region in INEL $pp$
collisions are analyzed in the present work. Over an energy range
from a few GeV to above 10 TeV, the chemical potentials of light
flavor particles (pion, kaon, and proton) and quarks (up, down,
and strange quarks) are extracted from the mentioned yield ratios
in which the contributions of strong decay from high-mass
resonance and weak decay from heavy flavor hadrons are removed.
Most energy dependent chemical potentials show the maximum at
about 4 GeV, while the energy dependent yield ratios do not show
such an extremum.
\\



\vskip1.0cm

\begin{multicols}{2}

{\section{Introduction}}

The yield ratios of negatively to positively charged pions
($\pi^-/\pi^+$), negatively to positively charged kaons
($K^-/K^+$), and anti-protons to protons ($\bar p/p$), as well as
the yield rations of other different particles are important
quantities measured in experiments, where the symbol of a given
particle is used for its yield for the purpose of simplicity.
Based on the yield ratios, one can obtain the chemical freeze-out
temperature ($T_{ch}$) of interacting system and the chemical
potential ($\mu_{baryon}$) of baryon in the framework of
statistical thermal model [1--4]. In the phase diagram of quantum
chromodynamics (QCD), $T_{ch}$ and $\mu_{baryon}$ describe
together the phase transition from hadronic matter to quark-gluon
plasma (QGP) or quark matter [4--7]. Except for $\mu_{baryon}$,
the chemical potentials of light particles (pion, kaon, and
proton) and light quarks (up, down, and strange quarks) are also
interesting and important in the studies of system evolution and
particle production.

According to the statistical thermal model [1--4], to study the
chemical potentials of light particles and quarks, we need the
yield ratios of $\pi^-/\pi^+$, $K^-/K^+$, and $\bar p/p$ at the
stage of chemical freeze-out at which inelastic collisions stop.
However, the data measured in experiments are usually at the stage
of past chemical freeze-out or kinetic freeze-out at which the
strong decay from high-mass resonance and weak decay from heavy
flavor hadrons contribute to the yield ratios [8], where the
kinetic freeze-out is a stage of system evolution at which the
probability density functions of particle momenta are invariant.
To use the expression of $T_{ch}$ and to obtain the chemical
potentials of light particles and quarks in the framework of
statistical thermal model [1--4], one should remove the
contributions of strong decay from high-mass resonance and weak
decay from heavy flavor hadrons to the yield ratios of
$\pi^-/\pi^+$, $K^-/K^+$, and $\bar p/p$ measured in experiments
[8].

Presently, the yield ratios of $\pi^-/\pi^+$, $K^-/K^+$, and $\bar
p/p$ produced in nucleus-nucleus and proton-proton ($pp$)
collisions at high energies are available to collect [9] in
experiments [6, 10--31]. Although the yield ratios in asymmetric
collisions are also available, we analyze more simply the yield
ratios in mid-rapidity interval in central gold-gold (Au-Au)
collisions at the Alternating Gradient Synchrotron (AGS) and the
Relativistic Heavy Ion Collider (RHIC) within its Beam Energy Scan
(BES) program, in central lead-lead (Pb-Pb) collisions at the
Super Proton Synchrotron (SPS) and the Relativistic Heavy Ion
Collider (RHIC), and in inelastic (INEL) or non-single-diffractive
(NSD) proton-proton ($pp$) collisions at the SPS and the Large
Hadron Collider (LHC), as well as in forward rapidity region in
INEL $pp$ collisions at the SPS at its BES. These data are
measured by some international collaborations over a
center-of-mass energy per nucleon pair ($\sqrt{s_{NN}}$) range
from a few GeV to above 10 TeV [6, 10--31].

In this paper, we analyze the chemical potentials of light
particles and quarks based on the yield ratios in the framework of
statistical thermal model [1--4]. Comparing with our recent work
[9], the contributions of strong decay from high-mass resonance
and weak decay from heavy flavor hadrons to the yield ratios are
removed. The energy dependent chemical potentials of light
particles and quarks are obtained.
\\

{\section{The method and formalism}}

To extract the chemical potentials of light particles and quarks,
the yield ratios of $\pi^-/\pi^+$, $K^-/K^+$, and $\bar p/p$
produced in Au-Au (Pb-Pb) and $pp$ collisions at the AGS, SPS at
its BES, RHIC at its BES, and LHC are needed, where the
contributions of strong and weak decays to the yield rations
should be removed. The same formula on the relation between the
yield ratio and chemical potential are used in our previous work
[9, 32] and the present work due to the standard and unified
expression. This results in some repetitions which are ineluctable
to give a whole representation of the present work.

In the framework of statistical thermal model of non-interacting
gas particles with the assumption of standard Boltzmann-Gibbs
statistics [1--4], based on the Boltzmann approximation in the
employ of grand-canonical ensemble, one has empirically [4, 5,
33--35]
\begin{equation}
T_{ch}=T_{\lim}\frac{1}{1+\exp\left[ 2.60-\ln\left( \sqrt{s_{NN}}
\right)/0.45\right]},
\end{equation}
where $\sqrt{s_{NN}}$ is in units of GeV and the ``limiting"
temperature $T_{\lim}\approx 0.16$ GeV. Meanwhile, based on the
Boltzmann approximation and the relation to isospin effect, one
has the relation among $\bar p/p$, $T_{ch}$, and chemical
potential $\mu_p$ of proton to be [17, 36, 37]
\begin{equation}
\frac{\bar{p}}{p} =\exp\left( -\frac{2\mu_{p}}{T_{ch}}\right)
\approx \exp\left(-\frac{2\mu_{baryon}}{T_{ch}}\right).
\end{equation}
Eqs. (1) and (2) are valid at the stage of chemical freeze-out
which is earlier than the strong decay from high-mass resonance
and weak decay from heavy flavor hadrons.

Similar to Eq. (2), $\pi^-/\pi^+$, $K^-/K^+$, and other two
negatively to positively charged particles ($D^-/D^+$ and
$B^-/B^+$) with together $\bar p/p$ are uniformly shown to be
\begin{align}
k_{j}&\equiv\frac{j^{-}}{j^{+}}
=\exp\left(-\frac{2\mu_{j}}{T_{ch}}\right),
\end{align}
where $j=\pi$, $K$, $p$, $D$, and $B$; $k_{j}$ denote the yield
ratio of negatively to positively charged particle $j$; and
$\mu_j$ denote the chemical potential of the particle $j$.

To obtain chemical potentials of quarks, the five yield ratios,
$k_{j}$ ($j=\pi$, $K$, $p$, $D$, and $B$), are enough. We shall
not discuss the yield ratio of top quark related antiparticles and
particles, top quark itself, and chemical potentials of top quark
related particle and top quark due to the fact that the lifetimes
of particles contained top quark are very short to be measured.

The chemical potential for quark flavor $q$ is denoted by
$\mu_{q}$, where $q=u$, $d$, $s$, $c$, and $b$ represent the up,
down, strange, charm, and bottom quarks, respectively. The values
of $\mu_{q}$ are then expected due to Eq. (3). According to refs.
[38, 39], $k_j$ ($j=\pi$, $K$, $p$, $D$, and $B$) are expressed by
$T_{ch}$ and $\mu_q$ ($q=u$, $d$, $s$, $c$, and $b$) to be
\begin{align}
k_{\pi} & =\exp\left[ -\frac{2\left(\mu_{u}-\mu_{d}\right)}
{T_{ch}}\right],\nonumber\\
k_{K} & =\exp\left[ -\frac{2\left( \mu_{u}-\mu_{s}\right)}
{T_{ch}}\right],\nonumber\\
k_{p} & =\exp\left[ -\frac{2\left(2\mu_{u}+\mu_{d}\right)}
{T_{ch}}\right],\nonumber\\
k_{D} & =\exp\left[ -\frac{2\left( \mu_{c}-\mu_{d}\right)}
{T_{ch}}\right],\nonumber\\
k_{B} & =\exp\left[ -\frac{2\left( \mu_{u}-\mu_{b}\right)}
{T_{ch}}\right].
\end{align}

According to Eqs. (3) and (4), $\mu_j$ of particle $j$ and $\mu_q$
of quark $q$ can be obtained in terms of $k_j$ or their
combination to be
\begin{align}
\mu_{j} & =-\frac{1}{2}T_{ch}\ln k_{j},
\end{align}
and
\begin{align}
\mu_{u} & =-\frac{1}{6}T_{ch}\left(\ln k_{\pi}+
\ln k_{p}\right),\nonumber\\
\mu_{d} & =-\frac{1}{6}T_{ch}\left( -2\ln k_{\pi}+\ln
k_{p}\right),\nonumber\\
\mu_{s} & =-\frac{1}{6}T_{ch}\left( \ln k_{\pi}-3\ln
k_{K}+\ln k_{p}\right),\nonumber\\
\mu_{c} & =-\frac{1}{6}T_{ch}\left( -2\ln k_{\pi}+\ln
k_{p}+3\ln k_{D}\right),\nonumber\\
\mu_{b} & =-\frac{1}{6}T_{ch}\left( \ln k_{\pi}+\ln k_{p}-3\ln
k_{B}\right),
\end{align}
respectively.

Although we show formula on $D$, $B$, $c$, and $b$ in Eqs.
(3)--(6), there is no $k_D$ and $k_B$ are analyzed in the present
work due to the limited data. The expressions on $D$, $B$, $c$,
and $b$ have only significance in methodology. In fact, the
present work focuses only $k_j$ and $\mu_j$ of light flavor
particles, $\pi$, $K$, and $p$, as well as $\mu_q$ of light flavor
quarks, $u$, $d$, and $s$.

It should be noted that Eq. (1) means a single-$T_{ch}$ scenario
for the chemical freeze-out. It is unambiguous that a two- or
multi-$T_{ch}$ scenario is also possible [40--44]. In the case of
using the two-$T_{ch}$, we need $T_{ch,S}$ for strange particles
and $T_{ch,NS}$ for non-strange particles. Thus, Eqs. (3)--(6) are
revised to
\begin{align}
k_K & \equiv \frac{K^-}{K^+}
=\exp\left(-\frac{2\mu_K}{T_{ch,S}}\right),\nonumber\\
k_j & \equiv \frac{j^-}{j^+}
=\exp\left(-\frac{2\mu_j}{T_{ch,NS}}\right), \hspace{2mm}
\left(j\neq K \right),
\end{align}
\begin{align}
k_{\pi} & =\exp\left[ -\frac{2\left(\mu_{u}-\mu_{d}\right)}
{T_{ch,NS}}\right],\nonumber\\
k_{K} & =\exp\left[ -\frac{2\left( \mu_{u}-\mu_{s}\right)}
{T_{ch,S}}\right],\nonumber\\
k_{p} & =\exp\left[ -\frac{2\left(2\mu_{u}+\mu_{d}\right)}
{T_{ch,NS}}\right],\nonumber\\
k_{D} & =\exp\left[ -\frac{2\left( \mu_{c}-\mu_{d}\right)}
{T_{ch,NS}}\right],\nonumber\\
k_{B} & =\exp\left[ -\frac{2\left( \mu_{u}-\mu_{b}\right)}
{T_{ch,NS}}\right],
\end{align}
\begin{align}
\mu_K & =-\frac{1}{2}T_{ch,S}\ln k_{j},\nonumber\\
\mu_j & =-\frac{1}{2}T_{ch,NS}\ln k_{j}, \hspace{2mm} \left(j\neq
K \right),
\end{align}
and
\begin{align}
\mu_{u} & =-\frac{1}{6}T_{ch,NS}\left( \ln k_{\pi} +
\ln k_{p}\right),\nonumber\\
\mu_{d} & =-\frac{1}{6}T_{ch,NS}\left(-2\ln k_{\pi} +
\ln k_{p}\right),\nonumber\\
\mu_{s} & =-\frac{1}{6}\left(T_{ch,NS}\ln k_{\pi}
- 3T_{ch,S}\ln k_{K} + T_{ch,NS}\ln k_{p}\right),\nonumber\\
\mu_{c} & =-\frac{1}{6}T_{ch,NS}\left(-2\ln k_{\pi} +
\ln k_{p} +3\ln k_{D}\right),\nonumber\\
\mu_{b} & =-\frac{1}{6}T_{ch,NS}\left(\ln k_{\pi} + \ln k_{p}
-3\ln k_{B}\right),
\end{align}
respectively.

The multi-$T_{ch}$ scenario will result in different chemical
freeze-out temperature $T_{ch,j}$ for emission of particles $j^-$
and $j^+$. In the case of considering the multi-$T_{ch}$ scenario,
Eqs. (3)--(6) should be revised to
\begin{align}
k_{j} \equiv \frac{j^{-}}{j^{+}}
=\exp\left(-\frac{2\mu_{j}}{T_{ch,j}}\right),
\end{align}
\begin{align}
k_{\pi} & =\exp\left[ -\frac{2\left(\mu_{u}-\mu_{d}\right)}
{T_{ch,\pi}}\right],\nonumber\\
k_{K} & =\exp\left[ -\frac{2\left( \mu_{u}-\mu_{s}\right)}
{T_{ch,K}}\right],\nonumber\\
k_{p} & =\exp\left[ -\frac{2\left(2\mu_{u}+\mu_{d}\right)}
{T_{ch,p}}\right],\nonumber\\
k_{D} & =\exp\left[ -\frac{2\left( \mu_{c}-\mu_{d}\right)}
{T_{ch,D}}\right],\nonumber\\
k_{B} & =\exp\left[ -\frac{2\left( \mu_{u}-\mu_{b}\right)}
{T_{ch,B}}\right],
\end{align}
\begin{align}
\mu_{j} =-\frac{1}{2}T_{ch,j}\ln k_{j},
\end{align}
and
\begin{align}
\mu_{u} & =-\frac{1}{6}\left( T_{ch,\pi}\ln k_{\pi} +
T_{ch,p}\ln k_{p}\right),\nonumber\\
\mu_{d} & =-\frac{1}{6}\left(-2T_{ch,\pi}\ln k_{\pi} +
T_{ch,p}\ln k_{p}\right),\nonumber\\
\mu_{s} & =-\frac{1}{6}\left(T_{ch,\pi}\ln k_{\pi}
- 3T_{ch,K}\ln k_{K} + T_{ch,p}\ln k_{p}\right),\nonumber\\
\mu_{c} & =-\frac{1}{6}\left(-2T_{ch,\pi}\ln k_{\pi} +
T_{ch,p}\ln k_{p} +3T_{ch,D}\ln k_{D}\right),\nonumber\\
\mu_{b} & =-\frac{1}{6}\left(T_{ch,\pi} \ln k_{\pi} + T_{ch,p}\ln
k_{p} -3T_{ch,B}\ln k_{B}\right),
\end{align}
respectively.

In the actual treatment in the present work, we shall use the
single-$T_{ch}$ scenario due to the fact that Eq. (1) is available
in literature [4, 5, 33, 34]. The two- or multi-$T_{ch}$ scenario
has only significance in methodology, though they are also
possible [40--44].
\\

{\section{Results and discussion}}

Figures 1(a), 1(b), and 1(c) present respectively the yield
ratios, $k_{\pi}$, $k_K$, and $k_p$, of negatively to positively
charged particles produced in mid-(pseudo)rapidity interval in
central Au-Au collisions, central Pb-Pb collisions, and INEL or
NSD $pp$ collisions, as well as in forward rapidity region in INEL
$pp$ collisions. The circles, squares, triangles, and stars
without $\bullet$, or the symbols with $+$ and without $\bullet$,
denote the yield ratios quoted in literature. The detailed
(pseudo)rapidity intervals, centrality ranges or collision types,
and collision systems are listed in Table 1 with together
collaborations and references. The circles, squares, triangles,
and stars with $\bullet$, or the symbols with $+$ and $\bullet$,
denote the yield ratios corrected to the primary production by
removing the contributions of strong decay from high-mass
resonance and weak decay from heavy flavor hadrons [8].

\begin{figure*}[!htb]
\begin{center}
\includegraphics[width=15.0cm]{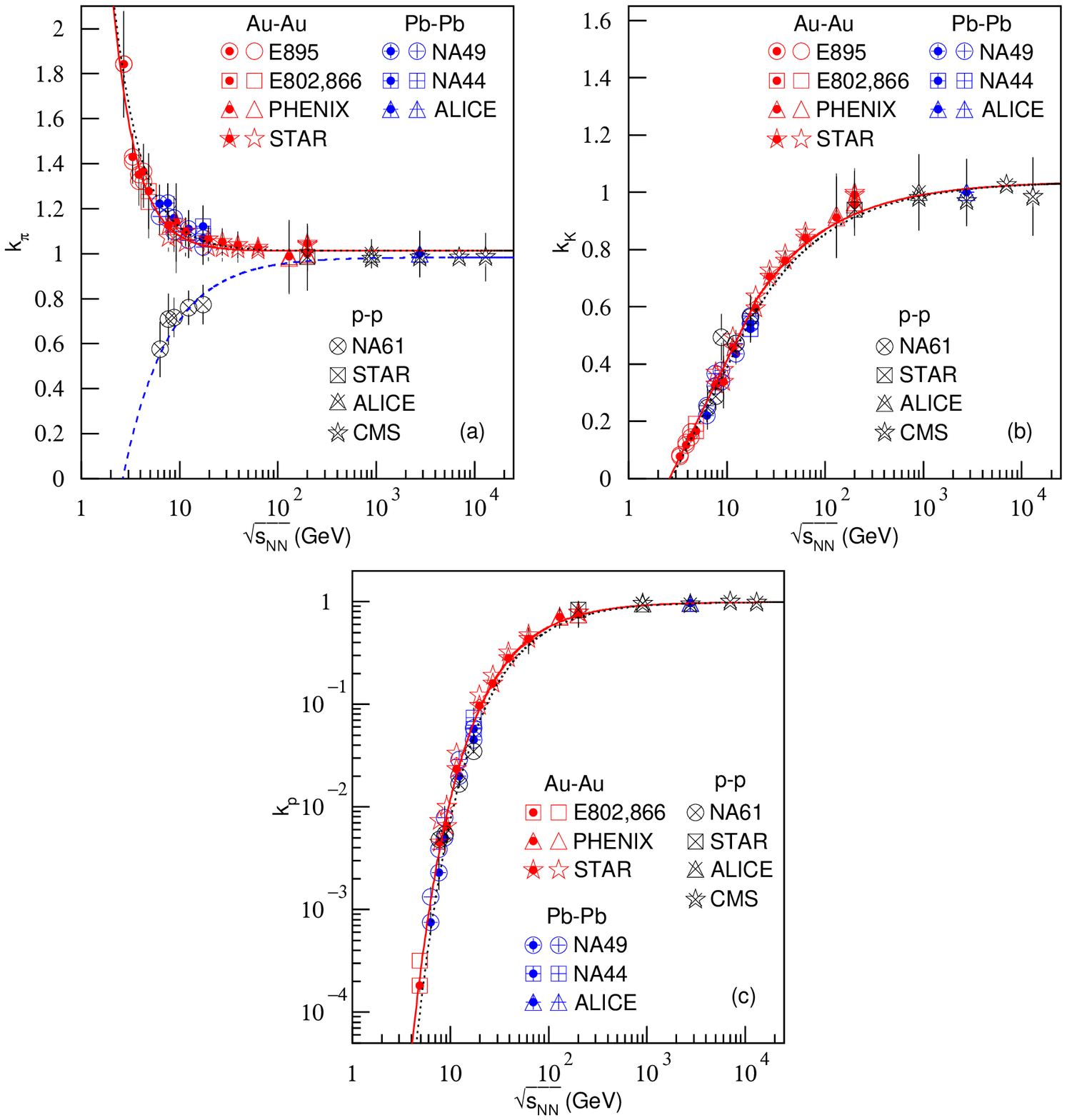}
\end{center}
Fig. 1. Yield ratios, (a) $k_{\pi}$, (b) $k_K$, and (c) $k_p$, of
negatively to positively charged particles produced in
mid-(pseudo)rapidity interval in central Au-Au collisions, central
Pb-Pb collisions, and INEL or NSD $pp$ collisions, as well as in
forward rapidity region in INEL $pp$ collisions. The circles,
squares, triangles, and stars without $\bullet$, or the symbols
with $+$ and without $\bullet$, denote the yield ratios quoted in
literature (see Table 1 for details). The circles, squares,
triangles, and stars with $\bullet$, or the symbols with $+$ and
$\bullet$, denote the yield ratios corrected to the primary
production by removing the contributions of strong decay from
high-mass resonance and weak decay from heavy flavor hadrons [8].
The curves are the results fitted by us for the $\sqrt{s_{NN}}$
dependent $k_j$ (see Eqs. (7)--(13) for details).
\end{figure*}

\begin{table*}[!htb]
{\scriptsize Table 1. The (pseudo)rapidity intervals, centrality
ranges or collision types, and collision systems corresponding to
the yield ratios quoted in Fig. 1.
\begin{center} \vskip-0.3cm
\begin{tabular}{cccccc}
\\ \hline\hline Open Symbol & (Pseudo)rapidity & Centrality or Type & Collisions & Collaboration & Reference \\
\hline
circles   & $|y|<0.05$ to $|y|<0.4$ & 0--5\%            & Au-Au, AGS  & E895, E866, E917 & [10--12] \\
squares   & $|y|<0.4$               & 0--10\%           & Au-Au, AGS  & E802, E866       & [13, 14] \\
triangles & $|\eta|<0.35$           & 0--5\%            & Au-Au, RHIC & PHENIX           & [15--17] \\
stars     & $|y|<0.1$ to $|y|<0.5$  & 0--5\% to 0--10\% & Au-Au, RHIC & STAR             & [6, 18--20] \\
circles with $+$        & $0<y<0.2$ or $|y|<0.1$ to $|y|<0.6$ & 0--5\% to 0--7.2\% & Pb-Pb, SPS & NA49       & [21--24] \\
squares with $+$        & $|y|<0.5$ to $|y|<0.85$             & 0--3.7\%           & Pb-Pb, SPS & NA44       & [25] \\
triangles with $+$      & $|y|<0.5$                           & 0--5\%             & Pb-Pb, LHC & ALICE      & [26] \\
circles with $\times$   & $y>0$                               & INEL               & $pp$, SPS  & NA61/SHINE & [27] \\
squares with $\times$   & $|y|<0.1$                           & NSD                & $pp$, RHIC & STAR       & [6, 28] \\
triangles with $\times$ & $|y|<0.5$                           & INEL               & $pp$, LHC  & ALICE      & [29] \\
stars with $\times$     & $|y|<1$                             & INEL               & $pp$, LHC  & CMS        & [30, 31] \\
\hline
\end{tabular}%
\end{center}}
\end{table*}

The solid (dotted) and dashed curves in Fig. 1(a) are the results
fitted by us for the $\sqrt{s_{NN}}$ dependent $k_{\pi}$ in
central Au-Au (Pb-Pb) collisions without (with) the corrections of
decays and in INEL or NSD $pp$ collisions respectively. The solid
(dotted) curves in Figs. 1(b) and 1(c) are the results fitted by
us for the $\sqrt{s_{NN}}$ dependent $k_K$ and $k_p$ respectively,
for the combining central Au-Au (Pb-Pb) collisions without (with)
the corrections of decays and INEL or NSD $pp$ collisions. One can
see that, with the increase of $\sqrt{s_{NN}}$, $k_{\pi}$
decreases obviously in central Au-Au (Pb-Pb) collisions and
increases obviously in INEL or NSD $pp$ collisions, and $k_K$ and
$k_p$ increase obviously in both central Au-Au (Pb-Pb) and INEL or
NSD $pp$ collisions.

The solid, dotted, and dashed curves in Fig. 1(a) can be
empirically described by
\begin{align}
k_{\pi}=&(4.212\pm0.682)\cdot(\sqrt{s_{NN}})^{-(1.799\pm0.152)} \nonumber \\
&+(1.012\pm0.019),
\end{align}
\begin{align}
k_{\pi}=&(3.712\pm0.611)\cdot(\sqrt{s_{NN}})^{-(1.519\pm0.148)} \nonumber \\
&+(1.012\pm0.019),
\end{align}
and
\begin{align}
k_{\pi}=&-(2.453\pm0.292)\cdot(\sqrt{s_{NN}})^{-(0.943\pm0.057)} \nonumber \\
&+(0.984\pm0.009),
\end{align}
respectively, with $\chi^2$/dof ($\chi^2$ per degree of freedom)
to be 0.162, 0.392, and 1.559 respectively. The solid and dotted
curves in Fig. 1(b) can be empirically described by
\begin{align}
k_K=&\big[-(0.291\pm0.028)+(0.306\pm0.010)\cdot\ln(\sqrt{s_{NN}})\big] \nonumber \\
&\cdot \theta(20-\sqrt{s_{NN}}) \nonumber \\
&+\big[-(2.172\pm0.146)\cdot(\sqrt{s_{NN}})^{-(0.554\pm0.018)} \nonumber \\
&\hskip4mm +(1.039\pm0.016)\big] \nonumber \\
&\cdot\theta(\sqrt{s_{NN}}-20)
\end{align}
and
\begin{align}
k_K=&\big[-(0.299\pm0.029)+(0.299\pm0.009)\cdot\ln(\sqrt{s_{NN}})\big] \nonumber \\
&\cdot \theta(20-\sqrt{s_{NN}}) \nonumber \\
&+\big[-(2.372\pm0.146)\cdot(\sqrt{s_{NN}})^{-(0.554\pm0.018)} \nonumber \\
&\hskip4mm +(1.039\pm0.016)\big] \nonumber \\
&\cdot\theta(\sqrt{s_{NN}}-20)
\end{align}
respectively, with $\chi^2$/dof to be 2.735 and 2.355
respectively. The solid and dotted curves in Fig. 1(c) can be
empirically described by
\begin{align}
k_p=&\exp\big[-(34.803\pm3.685)\cdot(\sqrt{s_{NN}})^{-(0.896\pm0.041)} \nonumber \\
&-(0.008\pm0.004)\big]
\end{align}
and
\begin{align}
k_p=&\exp\big[-(37.403\pm3.776)\cdot(\sqrt{s_{NN}})^{-(0.884\pm0.036)} \nonumber \\
&-(0.007\pm0.003)\big]
\end{align}
respectively, with $\chi^2$/dof to be 7.715 and 5.323
respectively.

The differences between the yield ratios without and with the
corrections of decays appear mainly over an energy range from a
few GeV to 100 GeV, though the differences are not very large. In
particular, the difference seems to be the largest at about 10
GeV. The limiting values of all the three yield ratios are one at
very high energy. According to the functions Eqs. (7)--(13), by
using Eqs. (5) and (6), the chemical potentials, $\mu_{\pi}$,
$\mu_K$, and $\mu_p$, of light particles, $\pi$, $K$, and $p$, as
well as the chemical potentials, $\mu_u$, $\mu_d$, and $\mu_s$, of
light quarks, $u$, $d$, and $s$, can be obtained respectively.

\begin{figure*}[!htb]
\begin{center}
\includegraphics[width=15.0cm]{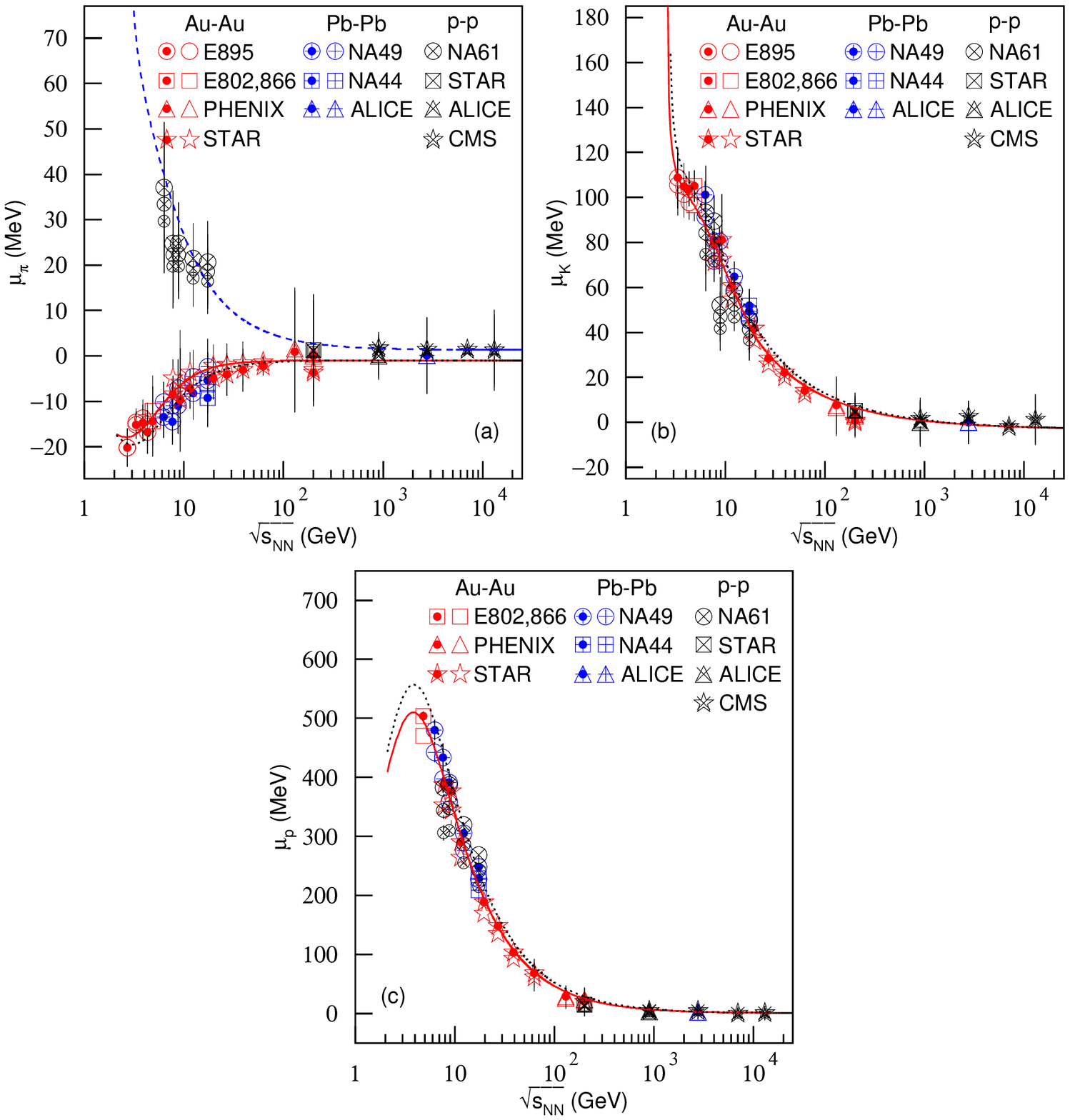}
\end{center}
Fig. 2. Chemical potentials, (a) $\mu_{\pi}$, (b) $\mu_K$, and (c)
$\mu_p$, of (a) $\pi$, (b) $K$, and (c) $p$ produced in
mid-(pseudo)rapidity interval in central Au-Au collisions, central
Pb-Pb collisions, and INEL or NSD $pp$ collisions, as well as in
forward rapidity region in INEL $pp$ collisions. The symbols
denote the derivative data obtained from Fig. 1 according to Eq.
(5). The normal, medium, and small symbols with diagonal crosses
denote the derivative data in INEL or NSD $pp$ collisions obtained
by $T_{ch}$, $0.9T_{ch}$, and $0.8T_{ch}$ in Eq. (5),
respectively. The curves surrounded the symbols are the derivative
results obtained from the curves in Fig. 1 according to Eq. (5).
\end{figure*}

\begin{figure*}[!htb]
\begin{center}
\includegraphics[width=15.0cm]{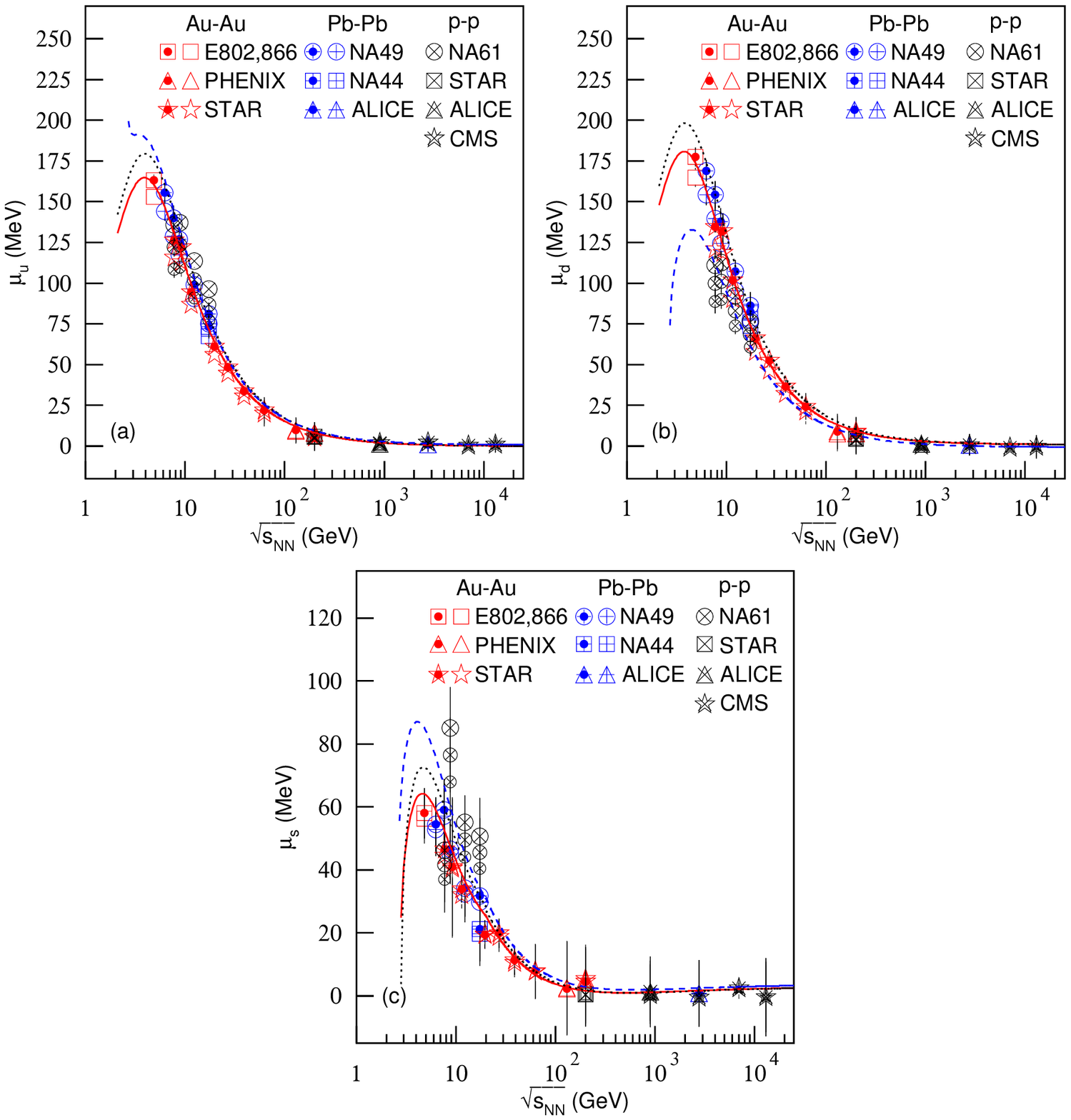}
\end{center}
Fig. 3. The same as Fig. 2, but showing the chemical potentials,
(a) $\mu_u$, (b) $\mu_d$, and (c) $\mu_s$, of (a) $u$, (b) $d$,
and (c) $s$ quarks according to Eq. (6). The solid (dotted) and
dashed curves are for central Au-Au (Pb-Pb) collisions without
(with) the corrections of decays and for INEL or NSD $pp$
collisions respectively.
\end{figure*}

The $\sqrt{s_{NN}}$ dependent $\mu_{\pi}$, $\mu_K$, and $\mu_p$
are shown in Figs. 2(a), 2(b), and 2(c), respectively. The symbols
denote the derivative data obtained from Fig. 1 according to Eq.
(5), where different symbols correspond to different
collaborations marked in the panels which are the same as Fig. 1.
Because of the chemical freeze-out temperature in $pp$ collisions
being unavailable, we use $T_{ch}$, $0.9T_{ch}$, and $0.8T_{ch}$
in Eq. (5) to obtain the derivative data in INEL or NSD $pp$
collisions, in which the corresponding results are orderly denoted
by normal, medium, and small symbols with diagonal crosses. One
can see that a low chemical freeze-out temperature in $pp$
collisions results in low chemical potentials.

In Fig. 2(a), the solid, dotted, and dashed curves represent the
same data samples as Fig. 1(a), but showing $\mu_{\pi}$. In Figs.
2(b) and 2(c), the solid and dotted curves represent the same data
samples as Figs. 1(b) and 1(c), but showing $\mu_K$ and $\mu_p$
respectively. One can see that, with the increase of
$\sqrt{s_{NN}}$, $\mu_{\pi}$ increases and decreases obviously in
central Au-Au (Pb-Pb) collisions and in INEL or NSD $pp$
collisions respectively, while $\mu_K$ and $\mu_p$ decrease
obviously in both central Au-Au (Pb-Pb) and INEL or NSD $pp$
collisions. At very high energy, all of $\mu_{\pi}$, $\mu_K$, and
$\mu_p$ approach to zero.

Figure 3 is the same as Fig. 2, but Figs. 3(a), 3(b), and 3(c)
present respectively the $\sqrt{s_{NN}}$ dependent $\mu_u$,
$\mu_d$, and $\mu_s$, which are derived from the symbols and
curves in Fig. 1 according to Eq. (6). The different symbols
correspond to different collaborations marked in the panels which
are the same as Figs. 1 and 2. The solid (dotted) and dashed
curves are for central Au-Au (Pb-Pb) collisions without (with) the
corrections of decays and for INEL or NSD $pp$ collisions
respectively, One can see that, with the increase of
$\sqrt{s_{NN}}$, $\mu_u$, $\mu_d$, and $\mu_s$ decrease obviously
in both central Au-Au (Pb-Pb) and INEL or NSD $pp$ collisions.
Like $\mu_{\pi}$, $\mu_K$, and $\mu_p$, all of $\mu_u$, $\mu_d$,
and $\mu_s$ also approach to zero at very high energy.

From Figs. 1--3 one can see that, in central Au-Au (Pb-Pb)
collisions, $k_{\pi}$ ($>1$) decreases obviously and $k_K$ ($<1$)
and $k_p$ ($<1$) increase obviously with the increase of
$\sqrt{s_{NN}}$. These differences also result in difference
between $\mu_{\pi}$ and $\mu_K$ ($\mu_p$). These differences are
caused by different mechanisms in productions of pions, kaons, and
protons. The contribution of strong and weak decays to $k_{\pi}$
is larger than those to $k_K$ and $k_p$. Comparing with pions,
kaons have larger cross-section of absorbtion in nuclei. In the
production of protons, the primary protons existed in the impact
nuclei also affect the yield.

At the top RHIC (200 GeV) and LHC energies, the trends of $k_j$,
$\mu_j$, and $\mu_q$ in central Au-Au (Pb-Pb) collisions are close
to those in INEL or NSD $pp$ collisions due to the increase of
hard scattering component. Finally, $k_j$ approaches to one and
$\mu_j$ and $\mu_q$ approaches to zero. These limiting values
render that the hard scattering process contributes largely, the
mean-free-path of produced particles (quarks) becomes largely, and
the viscous effect becomes weakly at the LHC. Meanwhile, the
interacting system changes completely from the hadron-dominant
state to the quark-dominant state at the early and medium stage of
collisions, though the final stage is hadron-dominant at the LHC.

The energy dependent $\mu_p$, $\mu_u$, $\mu_d$, and $\mu_s$ also
show the maximum at about 4 GeV, while the energy dependent
$\mu_{\pi}$, $\mu_K$, $k_{\pi}$, $k_K$, and $k_p$ do not show such
an extremum. The particular trend of the considered curves are
caused by some reasons. In terms of nuclear and hadronic
fragmentation, over an energy range from MeV to GeV, impact nuclei
undergone various modes of nuclear fission and fragmentation, as
well as multi-fragmentation and limiting fragmentation, then
hadronic fragmentation and limiting fragmentation appear. At the
stage of nuclear limiting fragmentation [45], nuclear fragments
have similar multiplicity and charge distributions. At the stage
of hadronic limiting fragmentation, the (pseudo)rapidity spectra
of relativistic produced particles in forward (backward) rapidity
region have the same or similar shape [46]. For heavy nucleus such
as Au and Pb, the initial energy of hadronic limiting
fragmentation is possibly about 4 GeV. In terms of phase
transition, about 4 GeV is possibly the initial energy of the
phase transition from a liquid-like state of nucleons and mesons
with a relatively short mean-free-path to a gas-like state of
nucleons and mesons with a relatively long mean-free-path in
central Au-Au (Pb-Pb) collisions.

Theoretically, chemical potentials always correspond to some
conserved charge. In Ref. [34], it is written how a hadron $j$ has
a chemical potential $\mu_j$. One has
\begin{align}
\mu_j=\mu_{baryon}B_j+\mu_S S_j+\mu_I I_j+\mu_C C_j,
\end{align}
where $B_j$, $S_j$, $I_j$, and $C_j$ are respectively the baryon
number, strangeness, isospin, and charm of the considered particle
$j$, and $\mu$ with lower foot marks $baryon$, $S$, $I$, and $C$
correspond to respective chemical potentials. Not all of the four
quantum numbers and four chemical potentials in the expression of
$\mu_j$ are free parameters since some of them are fixed by the
conservation laws and some of them are zero for a special
particle.

Both Eqs. (5) and (22) are obtained in the framework of
statistical thermal model [34, 36--39] or related literature [17].
These two formulas are different methods, but they should be
harmonious in description of particle chemical potential at the
stage of chemical freeze-out which is earlier than the strong and
weak decays. Using Eq. (5) with or without the corrections of
strong and weak decays causes a small difference of particle
chemical potentials. Using Eq. (22) we have concretely
$\mu_{\pi}=\mu_I I_{\pi}$, $\mu_K=\mu_S S_K+\mu_I I_K$, and
$\mu_p=\mu_{baryon}B_p+\mu_I I_p$ which should give similar
results to Eq. (5) with or without the corrections of strong and
weak decays. In particular, both Eqs. (5) and (22) results in zero
chemical potential at above top RHIC energy. However, Eq. (22) is
not available to determine $\mu_q$. Instead, the present work
shows a way to determine $\mu_j$ and $\mu_q$ simultaneously.

To determine $\mu_j$ for a given particle $j$ and $\mu_q$ for a
given quark $q$, the present work has used a simple, convenient,
and alternative method. In the case of utilizing $T_{ch}$, $\mu_j$
and $\mu_q$ can be obtained according to $k_j$ which is obtained
in experiments independently. Then, we can easily use Eq. (5) for
each particle independently and Eq. (6) for each quark
independently. In the extraction, we have neglected the difference
between the chemical potential $\mu_{j^-}$ of negatively charged
particle $j^-$ and the chemical potential $\mu_{j^+}$ of
positively charged particle $j^+$ due to small difference between
$\mu_{j^-}$ and $\mu_{j^+}$. Meanwhile, we have neglected the
difference between the chemical potential $\mu_{\bar q}$ of
anti-quark $\bar q$ and the chemical potential $\mu_q$ of quark
$q$ due to small difference between $\mu_{\bar q}$ and $\mu_q$.
Based on the above approximate treatment, Eqs. (1), (3), and (4)
are acceptable. Besides, we have used a single-$T_{ch}$ scenario
for the chemical freeze-out, though a two-$T_{ch}$ or
multi-$T_{ch}$ scenario is also possible.

Before summary and conclusions, it should be noted that although
the contributions of strong decay from high-mass resonance and
weak decay from heavy flavor hadrons [8] are excluded in the
present work, only one mode of decay affects mainly $k_{\pi}$,
$k_K$, or $k_p$ measured in experiments. For $k_{\pi}$, removing
the contribution of strong decay can regain the data from the
stage at primary production, where the strong decay pulls down
$k_{\pi}$. For $k_K$, removing the contribution of strong decay
can regain the data from the stage at primary production, where
the strong decay lifts $k_K$. For $k_p$, removing the weak decay
can regain the data from the stage at primary production, where
the weak decay lifts $k_p$. Generally, both strong and weak decays
do not affect largely the trends of experimental $k_j$ and then
$\mu_j$ and $\mu_q$, in particular at above top RHIC energy.

In the calculation on removing the contributions from strong and
weak decays from the data, we have utilized a very recent
literature [8] which works in the framework of statistical thermal
model [1--4]. In ref. [8], the energy dependent particle ratios
``from the stage at primary production, after strong decay from
high-mass resonance, and after weak decay from heavy flavor
hadrons" are presented. To compare with the data, the statistical
thermal model [1--4, 8] is coordinately accounted the effects of
experimental acceptance and transverse momentum cuts. What we do
in the present work is to directly quote the results obtained in
ref. [8]. One can see that strong decay affects mainly $k_{\pi}$
and $k_K$, while weak decay affects mainly $k_p$. Meanwhile, the
effect of quantum statistics is much smaller and can be neglected
[8].

In the case of including the contributions of two decays and
quantum statistics [8], the extracted energy dependent $\mu_j$ and
$\mu_q$ have small difference from those excluding the mentioned
contributions. Although the contributions of two decays to yields
of $\pi^-$ and $\pi^+$ are considerable, these effects to
$k_{\pi}$ are small. Except for the contributions to yields and
yield ratios, the two decays also contribute mainly in low
transverse momentum region and central rapidity interval. These
contributions affect more or less the trends of transverse
momentum and rapidity spectra in terms of slope or shape and
normalization constant. We shall not discuss the effects of two
decays on transverse momentum and rapidity spectra due to these
topics being beyond the focus of the present work.
\\

{\section{Summary and Conclusions}}

In summary, we have analyzed the yield ratios $k_{\pi}$, $k_K$,
and $k_p$ of negatively to positively charged particles produced
in mid-(pseudo)rapidity interval in central Au-Au collisions,
central Pb-Pb collisions, and INEL or NSD $pp$ collisions, as well
as in forward rapidity region in INEL $pp$ collisions over a
$\sqrt{s_{NN}}$ range from a few GeV to above 10 TeV. To obtain
the chemical potentials $\mu_j$ and $\mu_q$, $k_{\pi}$, $k_K$, and
$k_p$ are corrected by removing the contributions of strong decay
from high-mass resonance and weak decay from heavy flavor hadrons.
It is shown that, with the increase of $\sqrt{s_{NN}}$, $k_{\pi}$
($>1$) decreases obviously in central Au-Au (Pb-Pb) collisions,
$k_{\pi}$ ($<1$) increases obviously in INEL or NSD $pp$
collisions, and $k_K$ ($<1$) and $k_p$ ($<1$) increase obviously
in both central Au-Au (Pb-Pb) and INEL or NSD $pp$ collisions. The
limiting values of $k_{\pi}$, $k_K$, and $k_p$ are one at very
high energy.

The chemical potentials $\mu_{\pi}$, $\mu_K$, and $\mu_p$ of light
particles $\pi$, $K$, and $p$, as well as the chemical potentials
$\mu_u$, $\mu_d$, and $\mu_s$ of light quarks $u$, $d$, and $s$
are extracted from the corrected yield ratios in which there is no
contributions of two decays. With the increase of $\sqrt{s_{NN}}$
over a range from above a few GeV to above 10 TeV, $\mu_{\pi}$
($<0$) increases obviously in central Au-Au (Pb-Pb) collisions,
$\mu_{\pi}$ ($>0$) decreases obviously in INEL or NSD $pp$
collisions, and $\mu_K$ ($>0$) and $\mu_p$ ($>0$) decrease
obviously in both central Au-Au (Pb-Pb) and INEL or NSD $pp$
collisions. Meanwhile, $\mu_u$ ($>0$), $\mu_d$ ($>0$), and $\mu_s$
($>0$) decrease obviously in both central Au-Au (Pb-Pb) and INEL
or NSD $pp$ collisions. The limiting values of $\mu_{\pi}$,
$\mu_K$, $\mu_p$, $\mu_u$, $\mu_d$, and $\mu_s$ are zero at very
high energy. The difference between the results with and without
the correction of two decays is not too large.

Even though for that with the corrections of two decays, the same
particular energy is still existent as that without the
corrections. The energy dependent $\mu_p$, $\mu_u$, $\mu_d$, and
$\mu_s$ show the maximum at about 4 GeV, while the energy
dependent $\mu_{\pi}$, $\mu_K$, $k_{\pi}$, $k_K$, and $k_p$ do not
show such an extremum. For heavy nucleus such as Au and Pb, the
initial energy of limiting fragmentation is possibly about 4 GeV.
This energy is also possibly the initial energy of the phase
transition from a liquid-like state of nucleons and mesons with a
relatively short mean-free-path to a gas-like state of nucleons
and mesons with a relatively long mean-free-path in central Au-Au
(Pb-Pb) collisions. Meanwhile, the density of baryon number in
nucleus-nucleus collisions at this energy has a large value. These
particular factors render different trends of the considered
quantities at this energy.
\\
\\
{\bf Data availability}

The data used to support the findings of this study are included
within the article and are cited at relevant places within the
text as references.
\\
\\
{\bf Compliance with ethical standards}

The authors declare that they are in compliance with ethical
standards regarding the content of this paper.
\\
\\
{\bf Conflict of Interest}

The authors declare that they have no conflict of interest
regarding the publication of this paper. The funders had no role
in the design of the study; in the collection, analyses, or
interpretation of the data; in the writing of the manuscript, or
in the decision to publish the results.
\\
\\
{\bf Acknowledgments}

This work was supported by the National Natural Science Foundation
of China under Grant Nos. 11575103, 11847311, and 11747063, the
Scientific and Technological Innovation Programs of Higher
Education Institutions in Shanxi (STIP) under Grant No. 201802017,
the Shanxi Provincial Natural Science Foundation under Grant No.
201701D121005, the Fund for Shanxi ``1331 Project" Key Subjects
Construction, and the Doctoral Scientific Research Foundation of
Taiyuan University of Science and Technology under Grant No.
20152043.
\\

\end{multicols}

{\small
\begin{thebibliography}{99}
\setlength{\itemsep}{-1pt}

\bibitem {1}
J. Cleymans, B. K{\"a}mpfer, and S. Wheaton, Phys. Rev. C, {\bf
65}: 027901 (2002)
\bibitem {2}
F. Becattini, J. Manninen, and M. Ga{\'z}dzicki, Phys. Rev. C,
{\bf 73}: 044905 (2006)
\bibitem {3}
A. Andronic, P. Braun-Munzinger, K. Redlich, and J. Stachel, Nucl.
Phys. A, {\bf 789}: 334 (2007)
\bibitem {4}
J. Cleymans, H. Oeschler, K. Redlich, and S. Wheaton, Phys. Rev.
C, {\bf 73}: 034905 (2006)
\bibitem {5}
A. Andronic, P. Braun-Munzinger, and J. Stachel, Nucl. Phys. A,
{\bf 834}: 237c (2010)
\bibitem {6}
L. Adamczyk et al (STAR Collaboration), Phys. Rev. C, {\bf 96}:
044904 (2017)
\bibitem {7}
R. Bellwied, EPJ Web Conf., {\bf 171}: 02006 (2018)
\bibitem {8}
N. Yu and X. F. Luo, Eur. Phys. J. A, {\bf 55}: 26 (2019)
\bibitem {9}
H.-L. Lao, Y.-Q. Gao, and F.-H. Liu, Universe, {\bf 5}: 152 (2019)
\bibitem {10}
J. L. Klay et al (E895 Collaboration), Phys. Rev. C, {\bf 68}:
054905 (2003)
\bibitem {11}
L. Ahle et al (E866/E917 Collaboration), Phys. Lett. B, {\bf 490}:
53 (2000)
\bibitem {12}
J. L. Klay et al (E895 Collaboration), Phys. Rev. Lett., {\bf 88}:
102301 (2002)
\bibitem {13}
Y. Akiba for the E802 Collaboration, Nucl. Phys. A, {\bf 610}:
139c (1996)
\bibitem {14}
L. Ahle et al (E802 Collaboration), Phys. Rev. C, {\bf 57}: 466(R)
(1998)
\bibitem {15}
K. Adcox et al (PHENIX Collaboration), Phys. Rev. C, {\bf 69}:
024904 (2004)
\bibitem {16}
K. Adcox et al (PHENIX Collaboration), Phys. Rev. Lett., {\bf 88}:
242301 (2002)
\bibitem {17}
S. S. Adler et al (PHENIX Collaboration), Phys. Rev. C, {\bf 69}:
034909 (2004).
\bibitem {18}
B. Abelev et al (STAR Collaboration), Phys. Rev. C, {\bf 81}:
024911 (2010)
\bibitem {19}
B. Abelev et al (STAR Collaboration), Phys. Rev. C, {\bf 79}:
034909 (2009)
\bibitem{20}
J. Adams et al (STAR Collaboration), Phys. Rev. Lett., {\bf 92}:
112301 (2004)
\bibitem{21}
C. Alt et al (NA49 Collaboration), Phys. Rev. C, {\bf 77}: 024903
(2008)
\bibitem{22}
S. V. Afanasiev et al (NA49 Collaboration), Phys. Rev. C, {\bf
66}: 054902 (2002)
\bibitem{23}
C. Alt et al (NA49 Collaboration), Phys. Rev. C, {\bf 73}: 044910
(2006)
\bibitem{24}
S. V. Afanasiev et al (NA49 Collaboration), Phys. Rev. C, {\bf
69}: 024902 (2004)
\bibitem{25}
I. G. Bearden et al (NA44 Collaboration), Phys. Rev. C, {\bf 66}:
044907 (2002)
\bibitem{26}
B. Abelev et al (ALICE Collaboration), Phys. Rev. C, {\bf 88}:
044910 (2013)
\bibitem{27}
A. Aduszkiewicz et al (NA61/SHINE Collaboration), Eur. Phys. J. C,
{\bf 77}: 671 (2017)
\bibitem{28}
B. I. Abelev et al (STAR Collaboration), Phys. Rev. C, {\bf 75}:
064901 (2007)
\bibitem{29}
K. Aamodt et al (ALICE Collaboration), Eur. Phys. J. C, {\bf 71}:
1655 (2011)
\bibitem{30}
S. Chatrchyan et al (CMS Collaboration), Eur. Phys. J. C, {\bf
72}: 2164 (2012)
\bibitem{31}
A. M. Sirunyan et al (CMS Collaboration), Phys. Rev. D, {\bf 96}:
112003 (2017)
\bibitem{32}
Y.-Q. Gao, H.-L. Lao, and F.-H. Liu, Adv. High Energy Phys., {\bf
2018}: 6047960 (2018)
\bibitem {33}
A. Andronic, P. Braun-Munzinger, and J. Stachel, Acta Phys. Pol.
B, {\bf 40}: 1005 (2009)
\bibitem {34}
A. Andronic, P. Braun-Munzinger, and J. Stachel, Nucl. Phys. A,
{\bf 772}: 167 (2006)
\bibitem {35}
A. Andronic, P. Braun-Munzinger, K. Redlich, and J. Stachel,
Nature, {\bf 561}: 321 (2018)
\bibitem {36}
P. Koch, J. Rafelski, and W. Greiner, Phys. Lett. B, {\bf 123}:
151 (1983)
\bibitem {37}
P. Braun-Munzinger, D. Magestro, K. Redlich, and J. Stachel, Phys.
Lett. B {\bf 518}: 41 (2001)
\bibitem {38}
I. Arsene et al (BRAHMS Collaboration), Nucl. Phys. A, {\bf 757}:
1 (2005)
\bibitem {39}
H. Zhao and F.-H. Liu, Adv. High Energy Phys., {\bf 2015}: 137058
(2015)
\bibitem {40}
S. Chatterjee, S. Das, L. Kumar, D. Mishra, B. Mohanty, R. Sahoo,
and N. Sharma, Adv. High Energy Phys., {\bf 2015}: 349013 (2015)
\bibitem {41}
S. Chatterjee, B. Mohanty, and R. Singh, Phys. Rev. C, {\bf 92}:
024917 (2015)
\bibitem {42}
S. Chatterjee and B. Mohanty, Phys. Rev. C, {\bf 90}: 034908
(2014)
\bibitem {43}
D. Thakur, S. Tripathy, P. Garg, R. Sahoo, and J. Cleymans, Adv.
High Energy Phys., {\bf 2016}: 4149352 (2016)
\bibitem {44}
H.-L. Lao, H.-R. Wei, F.-H. Liu, and R. A. Lacey, Eur. Phys. J. A,
{\bf 52}: 203 (2016)
\bibitem {45}
O. Lindfors (edited), Introduction to Joint Institute for Nuclear
Research, Dubna, USSR, 1980, p.41 (unpublished)
\bibitem {46}
M. I. Adamovich et al (EMU01 Collaboration), Phys. Rev. Lett.,
{\bf 62}: 2801 (1989)

\end{thebibliography}}
\end{document}